\documentclass[aps,prb,twocolumn,superscriptaddress]{revtex4-1}
\usepackage{graphicx}
\usepackage{mathrsfs}
\usepackage{bm}
\usepackage{amsmath}
\usepackage{color}
\usepackage{dcolumn}
\usepackage{epstopdf}
\usepackage{dsfont}
\usepackage{amssymb}
\usepackage{tabularx}
\usepackage{array,ulem}
\usepackage{float}
\usepackage{colordvi}
\usepackage{verbatim}
\usepackage{hyperref}
\hypersetup{
	colorlinks=true,
	linkcolor=blue,
	filecolor=blue,
	urlcolor=blue,
	citecolor=blue,
}

\begin{document}
\title{
Two-Dimensional Higher-Order Topological Metals
}
\author{Lizhou Liu}
\affiliation{College of Physics, Hebei Normal University, Shijiazhuang 050024, China}

\author{Cheng-Ming Miao}
\affiliation{College of Physics, Hebei Normal University, Shijiazhuang 050024, China}
\affiliation{International Center for Quantum Materials, School of Physics, Peking University, Beijing 100871, China}

\author{Qing-Feng Sun}
\affiliation{International Center for Quantum Materials, School of Physics, Peking University, Beijing 100871, China}
\affiliation{Hefei National Laboratory, Hefei 230088, China}

\author{Ying-Tao Zhang}
\email[Correspondence author:~~]{zhangyt@mail.hebtu.edu.cn}
\affiliation{College of Physics, Hebei Normal University, Shijiazhuang 050024, China}

\date{\today}

\begin{abstract}
We investigate the energy band structure and energy levels of graphene with staggered intrinsic spin-orbit coupling and in-plane Zeeman fields. Our study demonstrates that staggered intrinsic spin-orbit coupling induces bulk band crossover at the the \( K \) and \( K' \) valleys and generates antihelical edge states at the zigzag boundaries, resulting in topological metallic phases. Quantized transport coefficients confirm the existence of these antihelical edge states. Furthermore, an in-plane Zeeman field, regardless of orientation, opens a gap in the antihelical edge states while preserving bulk band closure, leading to higher-order topological metals with corner states. We also validate the presence of these corner states in nanoflakes with zigzag boundaries and confirm the metallic phases with crossed bands through a continuum low-energy model analysis.
\end{abstract}

\maketitle
\section{Introduction}
In recent decades, topological states have flourished owing to their unique properties~\cite{Chiu2016, Bansil2016}. These states primarily encompass topological insulators~\cite{Hasan2010, LiuFeng}, topological superconductors~\cite{Qi2011, Sato2010}, topological semimetals~\cite{Armitage2018, Lv2021, Burkov2016, Dai2016}, and topological metal~\cite{Burkov2017, Bahari2019, Jangjan2021, Xie2021, Zuo2023, Yan2023}.
They all support topologically protected robust edge states or surface states, which are counted by the corresponding topological invariant~\cite{Thouless1982, Xiao2010, Teo2010}.
Depending on the bulk states, topological insulators and superconductors are characterized by gapped bulk states~\cite{Haldane1988, Kane2005, Kane2005a, Bernevig2006, Bernevig2006a}, and the energy band structure of topological semimetals is characterized by topologically protected band closures at the Fermi energy level.
Moreover, topological metals have conduction and valence bands intersecting at the Fermi level, leading to novel peculiar quasiparticle excitations~\cite{Ying2018, Ying2019, Xie2023, Dai2024}.

Recently, the concept of higher-order topological phase has been introduced~\cite{Benalcazar2017, Benalcazar2017a, Li2020, Benalcazar2019, Schindler2018, Peterson2018, Yao2023, Bhowmik2024, Liu2023, Hung2024, Mazanov2024}.
These systems exhibit zero-dimensional in-gap corner states (one-dimensional hinge states) in two-dimensional (three-dimensional) second-order topological states.
Initially, higher-order topological phases were proposed in spinless insulator systems ~\cite{Ezawa2018, Park2019, Liu2019, Sheng2019}.
Subsequently, by destroying the first-order edge states using in-plane Zeeman fields~\cite{Ren2020, Zhuang2022, Han2022, Miao2022, Miao2023, Miao2024, Chen2020a, Li2024} or interlayer coupling~\cite{Liu2024}, second-order topological insulators with topologically protected corner states have been widely studied.
Based on this, nontrivial second-order topological semimetal states have also been predicted in bulk-closed three-dimensional systems~\cite{Lin2018, Ghorashi2020, Wang2020b, Xiong2023, Pu2023, Ghorashi2021, Wang2022, Chen2022, Gao2023, Du2022, Hirsbrunner2024, Pan2024, Qi2024}.
However, realizing two-dimensional higher-order topological metals remains challenging due to the difficulty in gapping the edge states while preserving the metallic nature of the bulk.

\begin{figure}
  \centering
  \includegraphics[width=8.5cm,angle=0]{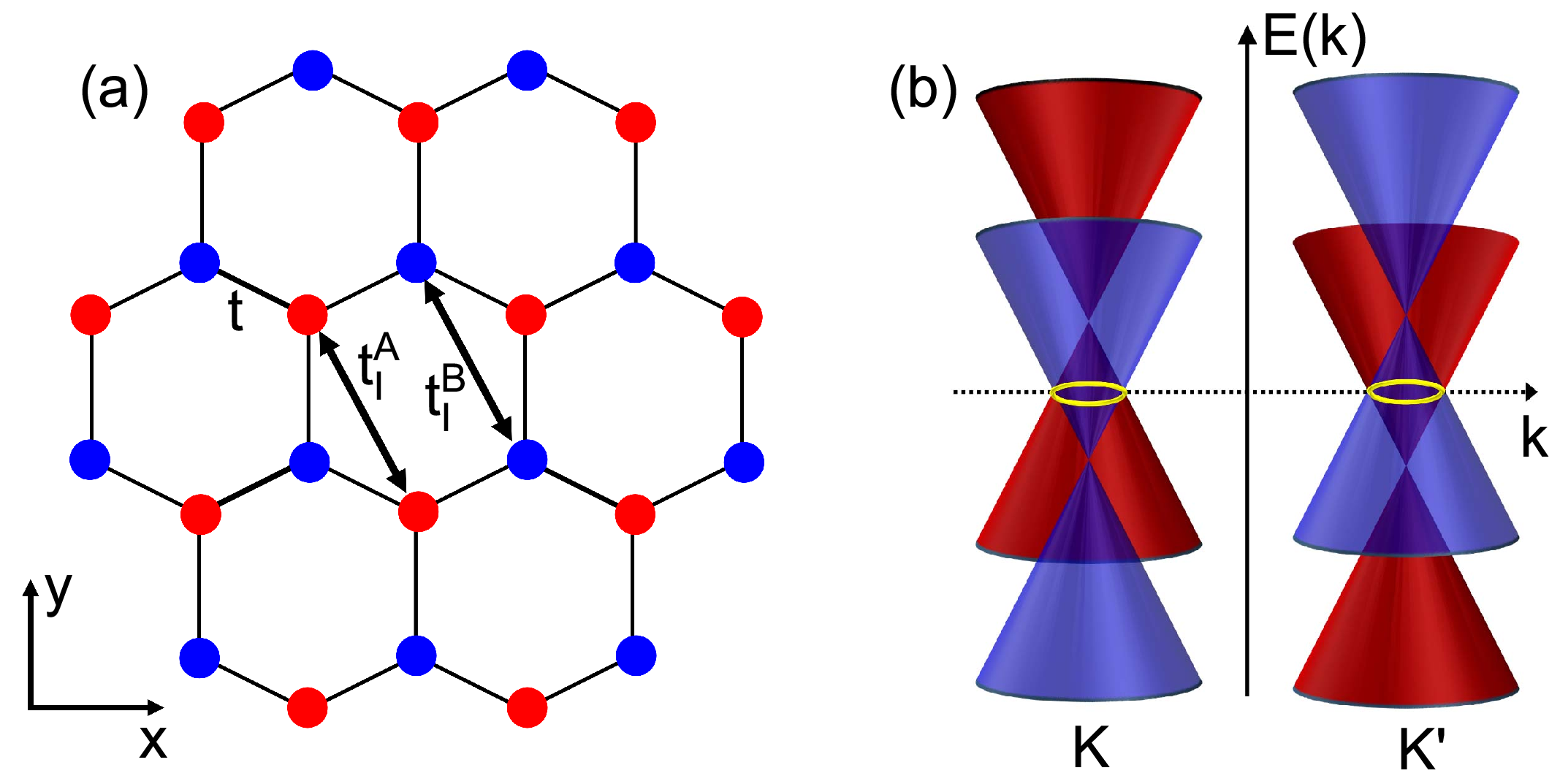}
\caption{(a) Schematic representation of the modified Kane-Mele model. Sublattices $A$ and $B$ are denoted by red and blue dots, respectively. The model includes spin-neutral nearest-neighbor hopping $t$ and spin-dependent sublattice-resolved next-nearest-neighbor intrinsic spin-orbit couplings $t_{\rm{I}}^{\rm A}$ and $t_{\rm{I}}^{\rm B}$. (b) Schematic illustration of bulk bands with staggered spin-orbit coupling. The red and blue cones represent the spin-up and spin-down components, respectively.}
  \label{fig1}
\end{figure}

In this work, we propose a scheme to realize two-dimensional second-order topological metal states using a modified Kane-Mele model.
As shown in Fig.~\ref{fig1}(a), the sublattices $A$ and $B$ of the modified Kane-Mele model are denoted by red and blue dots, respectively. The intrinsic spin-orbit couplings between sublattices $A$ and between sublattices $B$ have opposite signs, i.e., $t_{\rm{I}}^{\rm A} = -t_{\rm{I}}^{\rm B}$, respectively, which is feasible for graphene on transition-metal dichalcogenides substrates~\cite{Gmitra2015, Wang2015, Ghiasi2017, Zihlmann2018, Benitez2018}.
The modified Kane-Mele model is equivalent to two modified Haldane models with time-reversal symmetry~\cite{Colomes2018}. Consequently, the spin-up and spin-down bulk bands exhibit opposite energy shifts, as illustrated by the red and blue cones in Fig.~\ref{fig1}(b).
The relative displacement of the Weyl cones with opposite spins leads to the conduction band across the Fermi energy level and behaving as a metallic phase.
Moreover, antihelical edge states emerge between the $K$ and $K'$ valleys at zigzag boundaries. We confirm the presence of these antihelical edge states through quantized transmission coefficients obtained using the Landauer-B$\ddot{\rm u}$ttiker formalism. Introducing an in-plane Zeeman field results in gapped antihelical edge states at the zigzag boundary, leading to the emergence of in-gap corner states. These findings indicate the realization of two-dimensional higher-order topological metals with corner states.

\section{System Hamiltonian}

The tight-binding Hamiltonian of the modified Kane-Mele model is given by:
\begin{eqnarray}
H &=&-t\sum_{\langle ij \rangle}c^\dagger_{i}c_{j} + i t_{\rm{I}}^\gamma \sum_{\langle\langle ij \rangle\rangle}\nu_{ij}c^\dagger_{i}{s}_{z}c_{j}  +\lambda \sum_{i}c^{\dagger}_{i} \textbf{B}\cdot \textbf{s} c_{i}, \nonumber
\label{EQ1}
\end{eqnarray}
where $c^\dagger_{i} (c_{i})$ is the creation (annihilation) operator for an electron at site $i$, $\textbf{s}=(s_x, s_y, s_z)$ represent the Pauli matrices for spin.
The parameter $t$ corresponds to the nearest-neighbor hopping amplitude.
The $t_{\rm{I}}^{\gamma}$ describes the staggered intrinsic spin-orbit coupling involving next-nearest-neighbor hopping, with $\nu_{ij}=({\hat{\bm{d}}_i \times \hat{\bm{d}_j}})_z/{|\hat{\bm{d}}_i \times \hat{\bm{d}}_j|}$, where $\hat{\bm{d}}_{i,j}$ are unit vectors along the two bonds the particle traverses going from site $j$ to $i$.
The index $\gamma$ can be either $A$ or $B$, representing sublattice $A$ or $B$, with coupling strengths $t_{\rm{I}}^{\rm A} = -t_{\rm{I}}^{\rm B}$.
The final term represents the Zeeman field, oriented along the vector $\textbf{B}=(B_x, B_y)$ with a magnitude of $\lambda$. Unless otherwise specified, the Zeeman field is aligned in the x-direction.
Previous studies have shown that Rashba spin-orbit coupling leads to different spin expectation values in the conduction and valence bands, opening the valley bulk gap to destroy the metallic phase~\cite{Frank2018}, therefore Rashba spin-orbit coupling is not being considered.
Throughout this work, the Fermi level, intrinsic spin-orbit couplings, and Zeeman field are expressed in the unit of $t$.

To further analyze the model, we transform the Hamiltonian to momentum space. The momentum-space Hamiltonian can be expressed as follows:
\begin{eqnarray}
H(\textbf{k}) &=&[f_x(\textbf{k}) \sigma_x + f_y(\textbf{k}) \sigma_y ] s_0 + f_{\rm I} \sigma_0 s_z + \lambda \sigma_0 s_x,
\label{EQ2}
\end{eqnarray}
where
\begin{eqnarray}
f_x(\textbf{k}) &=&t [\cos a k_y +2\cos({ a k_y/2}) \cos({\sqrt{3}a k_x/2})], \nonumber  \\
f_y(\textbf{k}) &=& t[\sin a k_y- 2\sin({ a k_y/2}) \cos({\sqrt{3}a k_x/2})], \nonumber  \\
f_{\rm I}(\textbf{k}) &=&- 2 t_{\rm{I}}^{\rm A}[ \sin{\sqrt{3} a k_x} - 2 \cos({3 a k_y/2}) \sin({\sqrt{3} a k_x/2})], \nonumber
\label{EQ3}
\end{eqnarray}
with $\textbf{k}= (k_x, k_y)$ being quasi momentum and $a$ being the lattice constant.
The ${\bf \sigma}=(\sigma_x, \sigma_y, \sigma_z)$ are Pauli matrices for sublattice
($A$, $B$) space.  $\sigma_0$ and $s_0$ are the $2\times 2$ unit matrices for
the sublattice and spin spaces, respectively.

\begin{figure}
  \centering
  \includegraphics[width=8.5cm,angle=0]{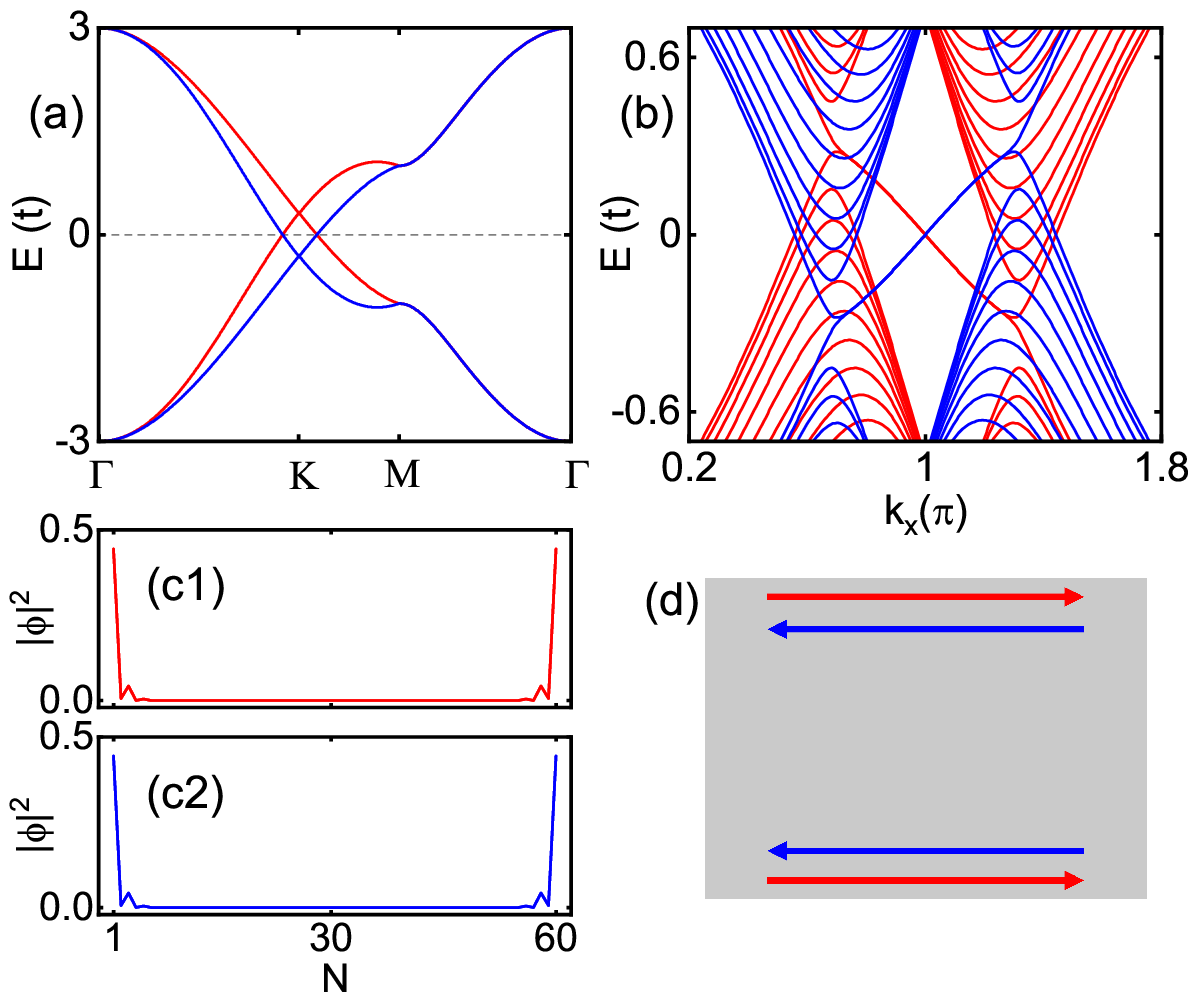}
 \caption{(a) Bulk energy spectrum of the modified Kane-Mele model.
 The spin-up (red lines) and spin-down (blue lines) bulk bands cross
 the Fermi energy level at the $K$ point.
 (b) Band structure of a zigzag nanoribbon with width $N_y = 60a$. Red and blue lines indicate spin-up and spin-down states, respectively.
 Twofold degenerate edge states connecting $K$ and $K'$ valleys.
 (c1) and (c2) show the probability distributions of spin-up and spin-down edge states
 for $k_x = 1.1 \pi$, respectively. (d) Schematic of edge state propagation in the zigzag edge geometry as shown in (b). The same (different) spin edge states propagate in the same (opposite) direction at two parallel boundaries, forming an antihelical edge state. Parameters are chosen as $t = 1$, $t_{\rm{I}}^{\rm A} = -t_{\rm{I}}^{\rm B} = 0.06$, $\lambda = 0$.}
  \label{fig2}
\end{figure}

\section{Antihelical edge state in metal phase}

To elucidate the properties of the staggered intrinsic spin-orbit coupling,
we analyze the electronic structure of the system in Fig.~\ref{fig2}. We initially investigate the bulk energy bands along the high-symmetry points with parameters $t = 1, t^{\rm A}_{\rm I}=-t^{\rm B}_{\rm I}=0.06$, and $\lambda = 0$, as shown in Fig.~\ref{fig2}(a).
In the absence of the Zeeman field, spin is a good quantum number, and the spin-up and spin-down energy bands are indicated by red and blue lines, respectively.
Remarkably, the spin-up valence band exceeds the lowest energy level of the spin-down conduction band at the $K$ point, and the energy parts of the two bands overlap each other [also see Fig.~\ref{fig1}(b)].
Thereby, the Fermi energy levels cross the two bands and the system exhibits a metallic phase.
Similarly, both the spin-down valence and spin-up conduction bands
at the $K'$ point passes through the Fermi energy also.
We then examine the edge states structure of zigzag nanoribbons in Fig.~\ref{fig2}(b), where the bulk energy bands intersect at the $K$ and $K'$ valleys. Two pairs of degenerate edge states emerge between the two valleys, with the red (blue) line representing spin-up (down) sectors.
The edge states with the same spin propagate in the same direction, whereas those with opposite spins propagate in opposite directions, characteristic of antihelical edge states~\cite{Xie2023}.
Furthermore, we present the spin-up and spin-down edge state distributions for $k_x = 1.1 \pi$ in Figs.~\ref{fig2}(c1) and~\ref{fig2}(c2), respectively.
The spin-up (spin-down) edge states are predominantly localized on the top and bottom sides of the zigzag nanoribbon.
The real-space distribution of these edge states is schematically illustrated in Fig.~\ref{fig2}(d), where the red and blue lines denote spin-up and spin-down edge states, respectively.
Consequently, the system exhibits the characteristics of a two-dimensional topological metal with antihelical edge states.
It should be emphasized that the $K$ and $K'$ valleys in graphene mix at the armchair boundary and are not good quantum numbers.
During the evolution from the Kane-Mele model to the modified Kane-Mele model, the valleys connecting the edge states are directly closed, leading to the disappearance of the edge states.

\begin{figure}
  \centering
  \includegraphics[width=8.5cm,angle=0]{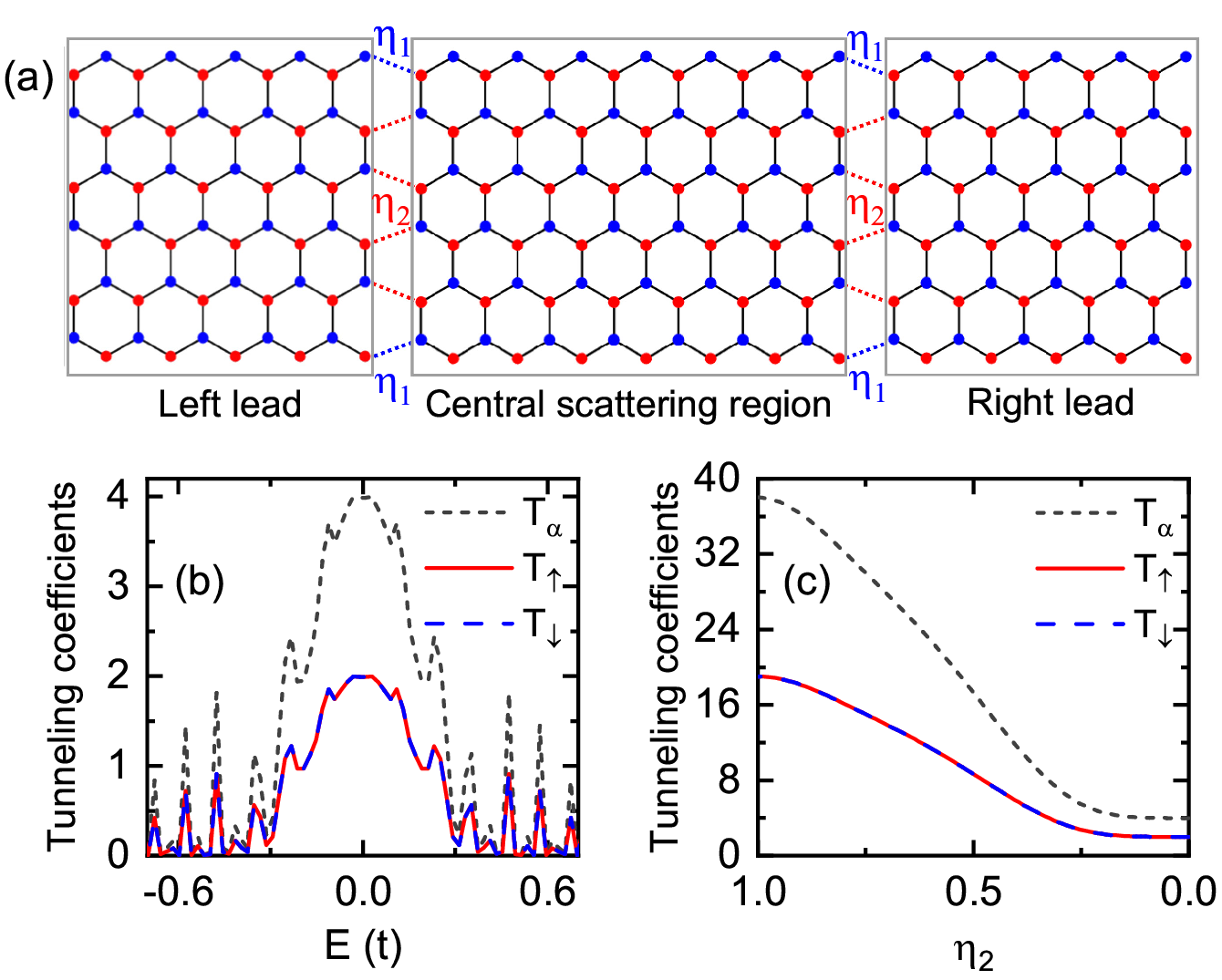}
\caption{(a) Schematics of the central scattering region connected by two semi-infinite leads, where the coupling strength $\eta_1$ ($\eta_2$) for the atoms at the edge (bulk) between the leads and the central scattering are indicated by the blue (red) dashed lines. Spin-resolved tunneling coefficients $T_{\uparrow/\downarrow}$ and total tunneling coefficient $T_{\alpha}$
versus (b) incident energy $E$ with $\eta_2=0$ and (c) bulk coupling strength $\eta_2$ with $E=0.01$. The tunneling coefficients $T_{\alpha}$,  $T_{\uparrow}$, and  $T_{\downarrow}$ are indicated by black dash line, red solid line, and blue dash line, respectively.
Other parameters are set to $t = 1$, $t_{\rm{I}}^{\rm A} = -t_{\rm{I}}^{\rm B} = 0.1$, $\lambda = 0$ and $\eta_1=1$.}
\label{fig3}
\end{figure}

The antihelical edge states in the modified Kane-Mele model can be experimentally detected using a two-terminal device, as illustrated in Fig. \ref{fig3}(a).
In this setup, $\eta_1$ ($\eta_2$) denotes the coupling strength for the atoms at the edge (bulk) between the leads and the central scattering region which can be controlled by the gate voltage.
We employ the nonequilibrium Greens function method to calculate the tunneling coefficient of the modified Kane-Mele model connected by two leads. According to the Landauer-B$\ddot{\rm u}$ttiker formula \cite{Fisher1981,Metalidis2005,Sun2009}, the differential conductance is given by $G_{\sigma}(E)=e^2/h \cdot T_{\sigma}(E)$, where the tunneling coefficient $T_{\sigma}(E)$ is defined as $T_{\sigma}(E)={\rm Tr}[{\rm \Gamma}_{L,\sigma} G^{r}_{\sigma} {\rm \Gamma}_{R,\sigma} (G^{r}_{\sigma})^{\dagger}]$. Here $\sigma$ represents the total spin $\alpha$, spin-up $(\uparrow)$ and spin-down $(\downarrow)$ components, respectively.
The linewidth functions for the left and right leads with spin $\sigma$ are ${\rm \Gamma}_{L,\sigma}(E)=i[\Sigma_{L,\sigma}^{r}-(\Sigma_{L,\sigma}^{r})^{\dagger}]$ and ${\rm \Gamma}_{R,\sigma}(E)=i[\Sigma_{R,\sigma}^{r}-(\Sigma_{R,\sigma}^{r})^{\dagger}]$. The retarded Green's function is given by $G^{r}_{\sigma}(E)=[E-H_{C,\sigma}-\Sigma_{L,\sigma}^{r}-\Sigma_{R,\sigma}^{r}]^{-1}$, where $H_{C,\sigma}$ is the Hamiltonian of the central scattering region.
The retarded self-energies terms contributed by the left and right leads are $\Sigma_{L,\sigma}^{r}=H_{CL,\sigma}g^{r}_{L,\sigma}H_{LC,\sigma}$ and $\Sigma_{R,\sigma}^{r}=H_{RL,\sigma}g^{r}_{R,\sigma}H_{RC,\sigma}$, where $g^{r}_{L,\sigma}$ and $g^{r}_{R,\sigma}$ are the surface Green's function of the semi-infinite leads \cite{Lee1981}. The coupling Hamiltonian between the central scattering region and lead is given by:
\begin{eqnarray}
H_{t}=&-t\eta_1\sum\limits_{\langle i'j' \rangle \in edge} c^\dagger_{i'}c_{j'} + i t_{\rm{I}}^\gamma \eta_1\sum\limits_{\langle\langle i'j' \rangle\rangle \in edge}\nu_{i'j'}c^\dagger_{i'}{s}_{z}c_{j'}  \nonumber
\\
&-t\eta_2\sum\limits_{\langle i'j' \rangle \in bulk} c^\dagger_{i'}c_{j'} + i t_{\rm{I}}^\gamma \eta_2\sum\limits_{\langle\langle i'j' \rangle\rangle \in bulk}\nu_{i'j'}c^\dagger_{i'}{s}_{z}c_{j'}. \nonumber
\\
\label{EQT}
\end{eqnarray}
Here, $H_{LC}=H_{CR}=H_{t}$ and $H_{CL}=H_{RC}=(H_{t})^{\dagger}$.

To intuitively demonstrate the existence of antihelical edge states, we plot the tunneling coefficients for different spins as a function of incident energy $E$ in Fig.~\ref{fig3}(b) with the bulk coupling strength $\eta_2=0$.
It is evident that a quantized platform $T_{\alpha}=2T_{\uparrow}=2T_{\downarrow}=4$ appears near $E=0$.
This indicates the existence of two parallel channels for single-spin electrons at the edge of the metallic system. In Fig. \ref{fig3}(c), we plot the tunneling coefficients for different spins as a function of the bulk coupling strength $\eta_2$ with the fixed incident energy $E=0.01$. Figure \ref{fig3}(c) shows that the tunneling coefficients $T_{\alpha}=2T_{\uparrow}=2T_{\downarrow}=38$ at $\eta _2=1$, indicating the participation of both edge and bulk states in transport. As the coupling strength $\eta_2$ decreases from 1 to 0, $T_{\alpha}$, $T_{\uparrow}$, and $T_{\downarrow}$ gradually decreases, due to the tunneling of bulk states being progressively suppressed.
As $\eta_2 < 0.2$, the bulk tunneling is completely suppressed, leaving only edge tunneling with $T_{\alpha}=2T_{\uparrow}=2T_{\downarrow}=4$.
These results demonstrate the presence of an antihelical edge state in the metallic phase of the modified Kane-Mele model.

\section{Corner states in metal phase}

\begin{figure}
  \centering
  \includegraphics[width=8.5cm,angle=0]{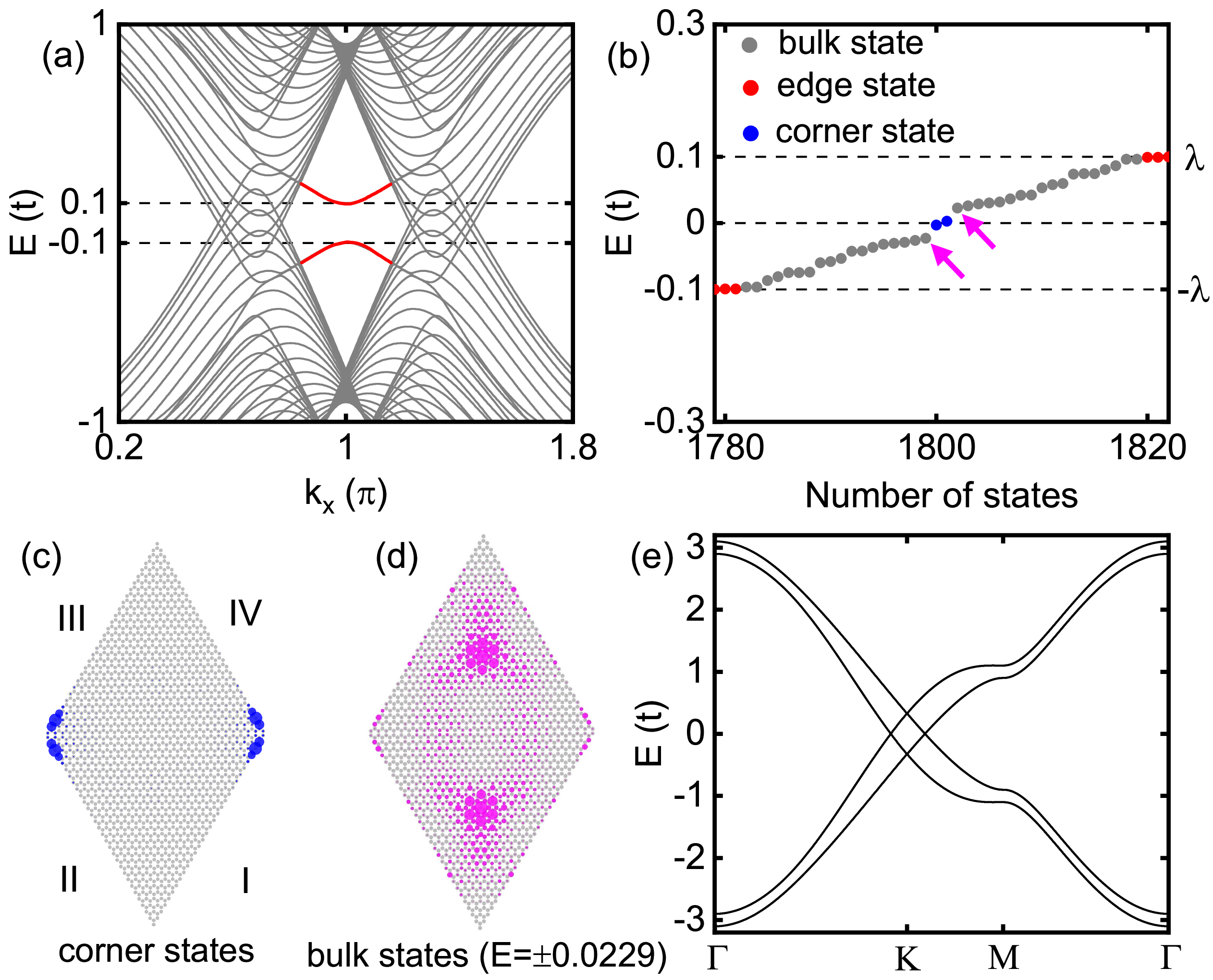}
\caption{(a) Band structures of zigzag ribbons of the modified Kane-Mele model with in-plane Zeeman field. The red curve indicates the gapped edge state. (b) Energy levels of the diamond-shaped nanoflake. Blue dots correspond to corner states, grey dots to bulk states, and red dots to edge states. (c) Probability distribution of the corner states (blue density dots). (d) Probability distribution of the bulk states (pink arrow pointing). (e) Bulk energy spectrum of the modified Kane-Mele model. Parameters are chosen as $t = 1$, $t_{\rm{I}}^{\rm A} = -t_{\rm{I}}^{\rm B} = 0.06$, $\lambda = 0.1$. The ribbon width $N_y = 60a$ and the nanoflake size is $60a \times 60a$.}
  \label{fig4}
\end{figure}

It is well known that an in-plane Zeeman field can break the time-reversal symmetry, thereby destroying the helical edge states in the Kane-Mele model and resulting in the appearance of corner states within the helical edge state gap \cite{Ren2020}. This raises the question: What is the effect of the in-plane Zeeman field on systems with the antihelical edge states that also satisfy time-reversal symmetry? To address this question, we introduce an in-plane Zeeman field to investigate its effect on the antihelical edge states and the metallic states.
In Fig.~\ref{fig4}(a), the energy band structure of the zigzag boundary nanoribbon is plotted, with an $x$-direction in-plane Zeeman field $\lambda = 0.1$, the other parameters being the same as in Fig.~\ref{fig2}.
Interestingly, the bulk energy band remains crossed at the $K$ and $K'$ valleys, whereas the antihelical edge states remain degenerate while become gapped, as indicated by the red line.
The staggered intrinsic spin-orbit coupling breaks the sublattice symmetry causing the bulk states crossing at the K and K' points, whereas the Zeeman field is independent of the sublattice and therefore maintains the bulk crossing.
The presence of gapped edge states in a two-dimensional system may suggest the emergence of higher-order corner states.

Now we turn our attention to validating the presence of the corner states. We employ a diamond-shaped nanoflake with zigzag boundaries of side length $60a$ and calculate the energy levels of the system, as shown in Fig.~\ref{fig4}(b).
Given the edge state gap of the system with zigzag nanoribbon is $2 \lambda$ shown in Fig.~\ref{fig4}(a), the edge state (red dot) gap in the energy level is also $2 \lambda$, as seen in Fig.~\ref{fig4}(b).
The bulk energy bands remain closed at the $K$ and $K'$ valleys represented by grey dots. Notably, two corner states (indicated by blue dots) appear around the zero Fermi energy.
In Fig.~\ref{fig4}(c), we present the wave-function probability distribution for these two corner states. Both states are uniformly distributed across both corners characteristic of higher-order topological phases.
To highlight the corner states, we plot the wave function distributions of neighboring bulk states (pink arrow pointing), as shown in Fig.~\ref{fig4}(d).
It is clearly seen that the bulk state is extended throughout the nanoflake, differing from the localized corner state.
It is worth noting that, due to the metallic nature, hybridization between the corner states and the bulk states is inevitable.
Consequently, the corner states exhibit a slight shift away from the zero energy as shown in Fig.~\ref{fig4}(b).
Additionally, a very small bulk gaps emerges at the zero energy in the nanoflake spectrum [see Fig.~\ref{fig4}(b)],
due to the size effect. This small bulk gap will become smaller as the nanodisk size increases, so it does not affect the metal properties.

For an in-depth understanding of the formation of corner states, we analytically investigate the effective mass of edge states in a modified Kane-Mele model subjected to an in-plane Zeeman field.
The four edges of the diamond-shaped nanoflake are labeled as $I$, $II$, $III$, and $IV$ in Fig.~\ref{fig4}(c).
The effective mass term ${\bf M}_{\lambda}^{\rm I}$ of edge $I$ can be achieved
\begin{eqnarray}
{{\bf M}_{\lambda}^I} &= & {\lambda}\frac{1-{{\rm{|}}\mu {\rm{|}}}}{{{\rm{|1 + }}\mu {\rm{|}}}}\left( {\begin{array}{*{20}{c}}
0&1\\
1&0
\end{array}} \right)  ={m_{\rm I}} \left( {\begin{array}{*{20}{c}}
0&1\\
1&0
\end{array}} \right),
\end{eqnarray}
with $\mu  =  {-1 + \frac{{{t^2}}}{{8t_{\rm I}^2}}}  + \sqrt {{{\left( {1 - \frac{{{t^2}}}{{8t_{\rm I}^2}}} \right)}^2} - 1}$.
Similarly, the effective mass term for the remaining three edges can be derived as follows: $m_{\rm II} = m_{\rm I}, m_{\rm III}=-m_{\rm I}$, and $m_{\rm {IV}} = -m_{\rm I}$, as explained in detail in Appendix~\ref{APPENDIX}.
 Since the absolute value of the Dirac effective mass at each zigzag boundary is equal, the gapped edge states induced by the Zeeman field remain degenerate [see Fig.~\ref{fig4}(a)].
Since the effective mass terms for edges $I$ and $II$ (as well as $III$ and $IV$) have the same sign, while the effective mass terms for $I$ and $IV$ (as well as $II$ and $III$) have opposite signs, Dirac mass domain walls are formed between $I$ and $IV$ (and $II$ and $III$).
According to the Jackiw-Rebbi mechanism~\cite{Jackiw1976}, these Dirac mass domain walls host topologically protected corner states [see Fig.~\ref{fig4}(c)].

Therefore, our results show that the staggered spin-orbit coupling breaks the sublattice symmetry leading to the metallic phase where the bulk band crosses.
In the absence of Zeeman field, the time reversal symmetry protects the existence of the antihelical edge state, and the bulk state compensates for the scattering of the co-directional edge state.
Once in-plane Zeeman field is introduced, the time-reversal symmetry is broken and there is a gap in the first-order edge states.
However, the Dirac mass domain wall between adjacent zigzag boundaries leads to the presence of corner states, forming higher-order topological metal phases.

The energy bands shown in Fig.~\ref{fig4}(a) confirm the persistence of the crossed band in the $K$ and $K'$ valleys. To better illustrate the effect of the in-plane Zeeman field on the bulk band, Fig.~\ref{fig4}(e) displays the bulk energy bands.
One can observe that the Zeeman field induces band splitting between the high-symmetry point $\Gamma - M$.
However, there is no observable splitting near the $K$ point, indicating that the metal phase remains intact.

To clearly understand and determine the band crossed of the metallic state, we employ the four-band low-energy continuum model expanded at valley $K$:
\begin{eqnarray}
H_{\rm eff}&=&3t/2 (\sigma_x k_x +\sigma_y k_y)s_0 +3\sqrt{3}t_{\rm I}\sigma_0 s_z +\lambda s_x \sigma_0.
\end{eqnarray}
where $t_{\rm I}=t_{\rm I}^A= -t_{\rm I}^B$.
The four eigenvalues satisfy the following relationship: $\varepsilon= \pm \sqrt{ \lambda^2 +27 t_{\rm I}^2 \pm 3 t \sqrt{(\lambda^2+ 27 t_{\rm I}^2 ) (k_x^2+ k_y^2) }+ \frac {9 k_x^2 t^2+ 9 k_y^2 t^2}{4}}.$
The low-energy effective Hamiltonian corresponds to the condition where the energy eigenvalues $\varepsilon=0$, leading to the following momentum relation:
\begin{eqnarray}
k_x^2+k_y^2=\frac{4 \lambda^2}{9 t^2}+\frac{12 t_{\rm I}^2}{t^2}.
\end{eqnarray}
This relationship indicates that the energy band cross at $\varepsilon=0$ form rings when in-plane Zeeman field $\lambda$ and staggered intrinsic spin-orbit coupling $t_{\rm I}$ take fixed values [see Fig.~\ref{fig1}(b)].
Therefore, the valence and conduction bands cross at the $K$ and $K'$ point Fermi energy level, ensuring a topological metallic phase.

For any in-plane Zeeman field direction, denoted as $\textbf{B}=(\cos{\theta}, \sin{ \theta})$, a spin rotation about the $z$-axis can be applied to align the Zeeman field along the $y$-direction. This is achieved using the unitary transformation matrix $\sigma_0 \otimes \exp[-i(\pi /2 - \theta)/2 s_z]$.
During this transformation, the terms in Eq.~(\ref{EQ2}) representing the lattice and spin-orbit coupling remain unchanged.
Thus, the Hamiltonians with an in-plane Zeeman field of any orientation can be continuously transformed into the $y$-direction.
This implies that two-dimensional higher-order metals can be induced with any direction in-plane Zeeman field.
If the magnetic field is out-of-plane, the antihelical edge states remain gapless despite the splitting caused by time-reversal symmetry breaking, and therefore does not induce higher-order topological phases.

\begin{figure}
  \centering
  \includegraphics[width=8.5cm,angle=0]{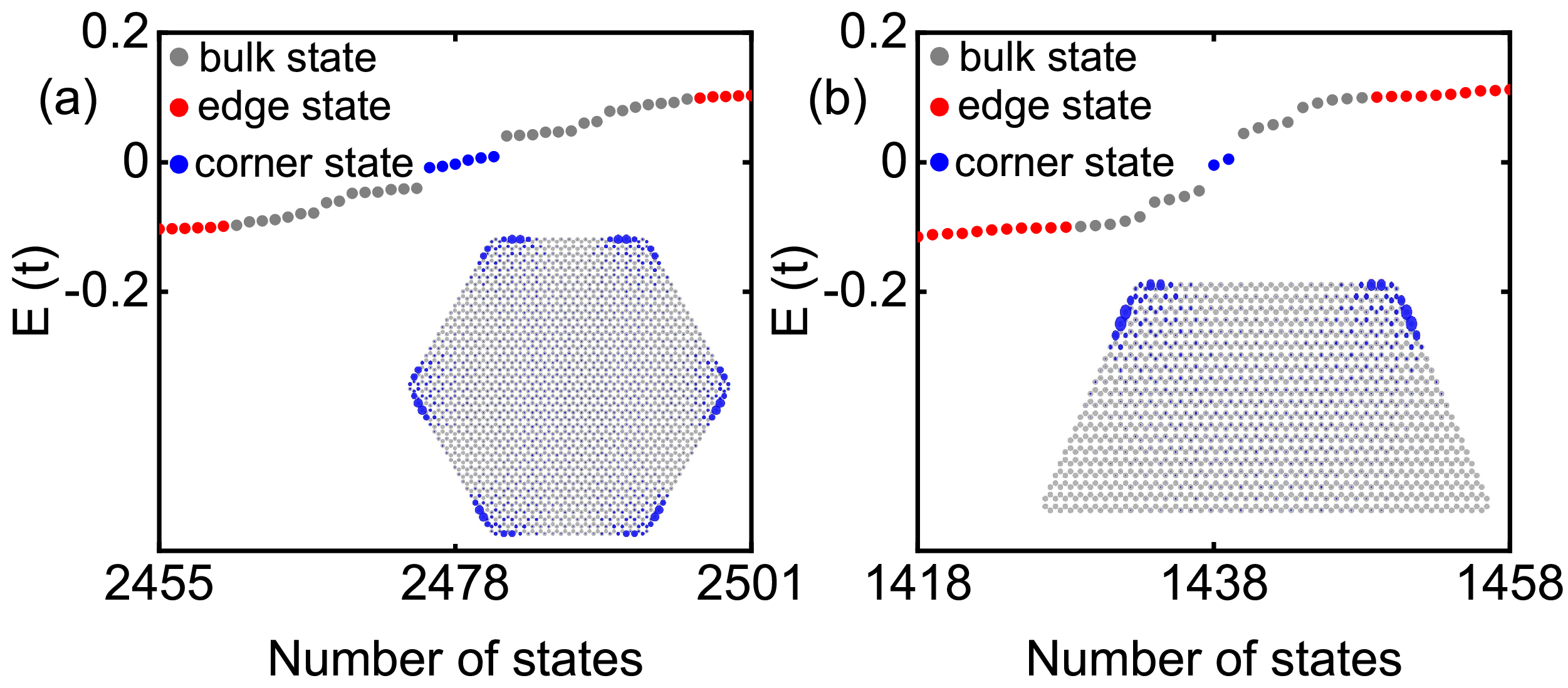}
\caption{Energy levels for (a) hexagon-shaped nanoflake and (b) trapezoid-shaped nanoflake with zigzag boundaries. Blue, grey, and red dots represent corner, bulk, and edge states, respectively. The inset shows the density distributions of the corner state denoted by a blue dot in the main figure.
Parameters are set to $t = 1$, $t_{\rm{I}}^{\rm A} = -t_{\rm{I}}^{\rm B} = 0.06$, $\lambda = 0.1$.}
  \label{fig5}
\end{figure}

\section{Corner states in hexagon-shaped and trapezoid-shaped nanoflakes}

Additionally, it is important to note that the topologically protected corner states are not limited to a specific nanoflake shape. In Fig.~\ref{fig5}, we calculate the energy levels and the wave function distributions of the corner states for hexagon-shaped and trapezoid-shaped nanoflakes with zigzag boundaries. As shown, the topologically protected corner states appear in every corner in hexagon-shaped nanoflakes, whereas in the trapezoid-shaped nanoflakes, these states localize only at the two obtuse angles.
To understand this behavior, we analyze the Dirac effective mass along the zigzag boundaries. The mass domain wall, induced by the in-plane Zeeman field, emerges specifically at the 120-degree angles. Consequently, the corner states appear at every corner in the hexagon-shaped nanoflake and the two obtuse angles in trapezoid-shaped nanoflakes. These observations demonstrate that the presence of these states is robust and can be observed in various geometric configurations, ensuring the universality of the two-dimensional second-order topological metal with corner states.

\section{Conclusions}

We have demonstrated that in-plane Zeeman fields can transform a modified Kane-Mele model into two-dimensional second-order topological metals with corner states.
Specifically, the staggered intrinsic spin-orbit coupling in the sublattice $A$ and $B$ shifts spin-up and spin-down Weyl cones in opposite energy directions, causing the valence and conduction bands to intersect at the Fermi energy level, thereby forming a topological metal.
The energy band structure and quantized transmission coefficients confirm the presence of antihelical edge states at the zigzag boundary.
An in-plane Zeeman field of arbitrary orientation opens a gap in these edge states.
To validate the presence of corner states, we examine a diamond-shaped nanoflake with zigzag boundaries.
Our calculations reveal two corner states in the energy spectrum, attributed to the effective mass domain walls at the corners. These states exhibit a uniform distribution over the two obtuse corners at half-filling. Additionally,
we also analyze the continuum low-energy model to confirm the topological metal phase.
Our findings provide compelling evidence for the realization of two-dimensional higher-order topological metals. We also point out that the strong staggered intrinsic spin-orbit coupling in graphene could be achieved by deposition on transition-metal dichalcogenides, such as $\rm {WS_2}$~\cite{Cummings2017, Benitez2018, Wang2015, Omar2017, Avsar2014}, $\rm {WSe_2}$~\cite{Volkl2017, Wang2016, Zihlmann2018}, $\rm {MoS_2}$~\cite{Benitez2018, Gmitra2015}, $\rm{MoSe_2}$~\cite{Ghiasi2017}.
In specific experiments, the methods for constructing graphene/transition-metal dichalcogenides heterojunctions include exfoliation and layers assembly~\cite{Britnell2013}, hydrothermal~\cite{Fang2015}, chemical vapor deposition~\cite{Li2018}, and solution-processing~\cite{Wang2017}.
By combining staggered spin-orbit coupling graphene with recently developed two-dimensional ferromagnets exhibiting in-plane anisotropy~\cite{Sheng2017, Liu2018, Webster2018, Cai2019, Umemoto2019, Bonilla2018, Jenjeti2018, Kim2018, Gong2019}, we are confident that our proposal model is both feasible and promising.

\section*{Acknowledgements}
This work was financially supported by the National Natural Science Foundation of China (Grants No. 12074097, No. 12374034, and No. 11921005),
Natural Science Foundation of Hebei Province (Grant No. A2024205025),
the Innovation Program for Quantum Science and Technology (Grant No. 2021ZD0302403),
and the Strategic Priority Research Program of the Chinese Academy of Sciences (Grant No. XDB28000000).

\appendix

\section{Analysis of antihelical edge states}\label{APPENDIX}

\begin{figure}[tbp]
  \centering
  \includegraphics[width=4.5cm,angle=0]{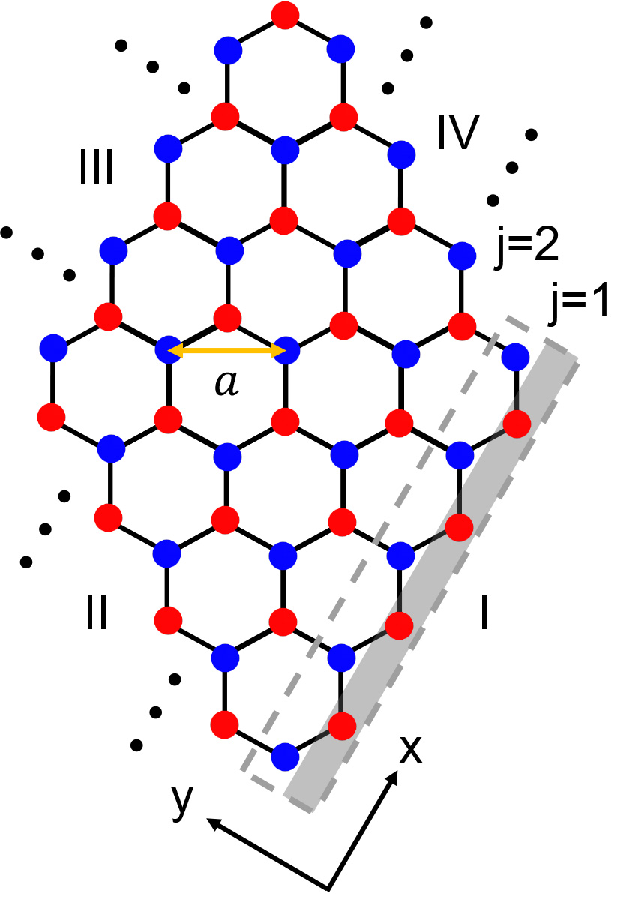}
  \caption{(a) The zigzag edge semi-infinite honeycomb lattice of a lattice constant $a$. Sublattices $A$ and $B$ are denoted by red and blue dots, respectively.
	The choice of superunit cells is illustrated by the gray dashed box.
	The corresponding colored square circled sublattices are the edge sublattice termination of the half-infinite sample area $y>0$.
	Here, $j$ is a real-space row index in the $y$ direction perpendicular to the edge.
 $I, II, III$ and $IV$ correspond to the boundary types of the diamond-shaped nanoflakes in Fig.~\ref{fig4}(c).
  }
  \label{figS1}
\end{figure}

We resolve the antihelical edge states by deriving the self-consistent equations for zigzag edge states in the modified Kane-Mele model semi-infinite lattice.
We consider a half-infinite sample area where $y>0$ with a zigzag boundary in the $x$-direction, as shown in Fig.~\ref{figS1}.
The tight-binding Hamiltonian without the Zeeman field can be expressed in momentum $q$ in the $x$-direction and the real space lattice index $j$ in the $y$-direction as:
\begin{align}
	&H(q)=\sum_{j}{\hat{\Psi}_{j}^{\dagger}(q) T_{0} (q)\hat{\Psi}_{j}(q)
	 +\hat{\Psi}_{j-1}^{\dagger}(q) T_{+1} (q)\hat{\Psi}_{j}(q)}\nonumber \\
	&\ \ \ \  \ \ \ \ +\hat{\Psi}_{j+1}^{\dagger}(q) T_{-1} (q)\hat{\Psi}_{j}(q),\nonumber \\
	&T_{0} (q)=2 t_{\rm{I}}^{\rm A} \sin {q} s_z \sigma_0+2 t \cos \frac{q}{2} s_0 \sigma_x,\nonumber \\ &T_{+1} (q)=-2 t_{\rm{I}}^{\rm A} \sin \frac{q}{2} s_z \sigma_0+t s_0 \left(\frac{\sigma_x-i \sigma_y}{2}\right),\nonumber \\
	&T_{-1}(q)=-2 t_{\rm{I}}^{\rm A} \sin \frac{q}{2} s_z  \sigma_0+t s_0 \left(\frac{\sigma_x+i \sigma_y}{2}\right).
	\label{eq2}
\end{align}
The staggered intrinsic spin-orbit coupling $t_{\rm{I}}^{\rm A}= - t_{\rm{I}}^{\rm B}$.
The basis vector is chosen as $\hat{\Psi}_{j}(q)
=(\hat{\Psi}_{j,A,\uparrow}(q),\hat{\Psi}_{j,B,\uparrow}(q),
\hat{\Psi}_{j,A,\downarrow}(q), \hat{\Psi}_{j,B,\downarrow}(q))^{T} $.
We construct Harper's equation \cite{Harper1955} which describes the wave function in real space in the direction normal to the edge as:
\begin{eqnarray}
&E(q)\Psi_{j}(q)=  T_{0}(q)\Psi_{j}(q)+ T_{+1}(q)\Psi_{j+1}(q)\nonumber \\
&+ T_{-1}(q)\Psi_{j-1}(q),
\label{eq3}
\end{eqnarray}
with the wave function $\Psi_{j}(q)$.
The $j$ dependence of the wave function can be simplified by setting $\Psi_{j}(q)=e^{\kappa(j-1)}\Psi_{1}(q)$.
So the relation between the wave function of the $j+1$ ($j-1$) row and the $j$ row is given by $\Psi_{j+1}(q)=e^{\kappa}\Psi_{j}(q)$ ($\Psi_{j-1}(q)=e^{-\kappa}\Psi_{j}(q)$).
We can rewrite Eq. (\ref{eq3}) as follow:
\begin{eqnarray}
E(q)\Psi_{j}(q)&=& [T_{0}(q)+ T_{+1}(q)e^{\kappa}+ T_{-1}(q)e^{-\kappa}] \Psi_{j}(q)\nonumber \\
    &=&\mathcal{H}^{\rm I} \Psi_{j}(q),
\label{eq}
\end{eqnarray}
where $\mathcal{H}^{\rm I}$ is the effective Hamiltonian of edge $I$. Utilizing Eqs. (\ref{eq2}) and (\ref{eq}), the effective Hamiltonian of edge $I$ can be written as:
\begin{eqnarray}
	&\mathcal{H}^{\rm I} &=T_{0}(q)+ e^{\kappa}T_{+1}(q)+e^{-\kappa}T_{-1}(q) \nonumber \\
	&&=4 t_{\rm{I}}^{\rm A} \sin \frac{q}{2}\left(\cos \frac{q}{2}-\cosh \kappa\right) s_z\sigma_0 - i t \sinh \kappa s_0\sigma_y\nonumber \\
	&&+\left(2 t \cos \frac{q}{2}+t \cosh \kappa\right) s_0\sigma_x.
	\label{eq4}
\end{eqnarray}
The antihelical gapless edge states cross at the time-reversal invariant point $q=\pi$ with energy $\varepsilon=0$.
It is worth emphasizing that the semi-infinite sample area where $y>0$, the outermost sublattices are all $A$, corresponds to the boundary $I$ and $II$ of the diamond-shaped nanoflake, as shown in Fig.~\ref{figS1}.
In this condition, the eigenstates of the spin-decoupled Hamiltonian at boundaries $I$ and $II$ in the diamond-shaped nanoflake can be represented by the following spinors:
\begin{eqnarray}
{\left| {{\chi _ {\uparrow} }} \right\rangle ^{\rm {I/II}}} &=
&\frac{1}{{\sqrt {1 + \mu } }} {\left[ {\begin{array}{*{20}{c}}
1\\
{ - i\sqrt \mu  }
\end{array}} \right]_\sigma } \otimes {\left[ {\begin{array}{*{20}{l}}
1\\
0
\end{array}} \right]_s}, \nonumber\\ 
{\left| {{\chi _ {\downarrow} }} \right\rangle ^{\rm {I/II}}} &=
&\frac{1}{{\sqrt {1 + \mu } }} {\left[ {\begin{array}{*{20}{c}}
1\\
{ i\sqrt \mu  }
\end{array}} \right]_\sigma } \otimes {\left[ {\begin{array}{*{20}{l}}
0\\
1
\end{array}} \right]_s},  \nonumber \\
\end{eqnarray}
where $\mu  =  {-1 + \frac{{{t^2}}}{{8t_{\rm I}^2}}}  + \sqrt {{{\left( {1 - \frac{{{t^2}}}{{8t_{\rm I}^2}}} \right)}^2} - 1}$.

However, when $y<0$, the outermost sublattice of the edge $I$ is $B$, which corresponds to the boundaries $III$ and $IV$ of the diamond-shaped nanoflake, as shown in Fig.~\ref{figS1}.
In the same way, the eigenstates of the spin-decoupled edge Hamiltonian can be represented by the following spinors:
\begin{eqnarray}
{\left| {{\chi _ {\uparrow} }} \right\rangle ^{\rm {III/IV}}} &=
&\frac{1}{{\sqrt {1 + 1/\mu } }} {\left[ {\begin{array}{*{20}{c}}
1\\
{ - \frac{i}{\sqrt \mu}  }
\end{array}} \right]_\sigma } \otimes {\left[ {\begin{array}{*{20}{l}}
1\\
0
\end{array}} \right]_s}, \nonumber\\ 
{\left| {{\chi _ {\downarrow} }} \right\rangle ^{\rm {III/IV}}} &=
&\frac{1}{{\sqrt {1 + 1/\mu } }}{\left[ {\begin{array}{*{20}{c}}
1\\
{ \frac{i}{\sqrt \mu}  }
\end{array}} \right]_\sigma } \otimes {\left[ {\begin{array}{*{20}{l}}
0\\
1
\end{array}} \right]_s}. 
\end{eqnarray}

The effective mass term ${\bf M}_{\lambda}^{\rm I}$ of the edge $I$ can be obtained by projecting the in-plane Zeeman field term $H_\lambda= \lambda {\sigma _0} {s_x}$ onto the subspace spanned by $\left| {\chi _ {\uparrow} ^I} \right\rangle$, $\left| {\chi _ {\downarrow} ^I} \right\rangle$.
\begin{eqnarray}
{{\bf M}_{\lambda}^I} &= & \lambda \left( {\begin{array}{*{20}{c}}
{\langle \chi _ {\uparrow} ^I | {\sigma _0}{s_x} |  {\chi _ {\uparrow} ^I} \rangle }&{\langle \chi _ {\uparrow} ^I | {\sigma _0}{s_x} |  {\chi _ {\downarrow} ^I} \rangle }\\
{\langle \chi _ {\downarrow} ^I | {\sigma _0}{s_x} |  {\chi _ {\uparrow} ^I}   \rangle }&{\langle \chi _ {\downarrow} ^I | {\sigma _0}{s_x} |  {\chi _ {\downarrow} ^I} \rangle }
\end{array}} \right)\nonumber\\
& = & {\lambda}\frac{1-{{\rm{|}}\mu {\rm{|}}}}{{{\rm{|1 + }}\mu {\rm{|}}}}\left( {\begin{array}{*{20}{c}}
0&1\\
1&0
\end{array}} \right)  ={m_{\rm I}} \left( {\begin{array}{*{20}{c}}
0&1\\
1&0
\end{array}} \right).
\end{eqnarray}
Similarly, the effective mass term of the remaining three edges can be obtained
\begin{eqnarray}
{{\bf M}_{\lambda}^{\rm II}} &= & \lambda \left( {\begin{array}{*{20}{c}}
{\langle \chi _ {\uparrow} ^{\rm II} | {\sigma _0}{s_x} |  {\chi _ {\uparrow} ^{\rm II}} \rangle }&{\langle \chi _ {\uparrow} ^{\rm II} | {\sigma _0}{s_x} |  {\chi _ {\downarrow} ^{\rm II}} \rangle }\\
{\langle \chi _ {\downarrow} ^{\rm II} | {\sigma _0}{s_x} |  {\chi _ {\uparrow} ^{\rm II}}   \rangle }&{\langle \chi _ {\downarrow} ^{\rm II} | {\sigma _0}{s_x} |  {\chi _ {\downarrow} ^{\rm II}} \rangle }
\end{array}} \right)\nonumber\\
& = & {\lambda}\frac{1-{{\rm{|}}\mu {\rm{|}}}}{{{\rm{|1 + }}\mu {\rm{|}}}}\left( {\begin{array}{*{20}{c}}
0&1\\
1&0
\end{array}} \right)  ={m_{\rm II}} \left( {\begin{array}{*{20}{c}}
0&1\\
1&0
\end{array}} \right),\nonumber\\ 
{{\bf M}_{\lambda}^{\rm III}} &= & \lambda \left( {\begin{array}{*{20}{c}}
{\langle \chi _ {\uparrow} ^{\rm III} | {\sigma _0}{s_x} |  {\chi _ {\uparrow} ^{\rm III}} \rangle }&{\langle \chi _ {\uparrow} ^{\rm III} | {\sigma _0}{s_x} |  {\chi _ {\downarrow} ^{\rm III}} \rangle }\\
{\langle \chi _ {\downarrow} ^{\rm III} | {\sigma _0}{s_x} |  {\chi _ {\uparrow} ^{\rm III}}   \rangle }&{\langle \chi _ {\downarrow} ^{\rm III} | {\sigma _0}{s_x} |  {\chi _ {\downarrow} ^{\rm III}} \rangle }
\end{array}} \right)\nonumber\\
& = & {\lambda}\frac{{{\rm{|}}\mu {\rm{|}}} -1}  {{{\rm{|1 + }}\mu {\rm{|}}}}\left( {\begin{array}{*{20}{c}}
0&1\\
1&0
\end{array}} \right)  ={m_{\rm III}} \left( {\begin{array}{*{20}{c}}
0&1\\
1&0
\end{array}} \right),\nonumber\\ 
{{\bf M}_{\lambda}^{\rm IV}} &= & \lambda \left( {\begin{array}{*{20}{c}}
{\langle \chi _ {\uparrow} ^{\rm IV} | {\sigma _0}{s_x} |  {\chi _ {\uparrow} ^{\rm IV}} \rangle }&{\langle \chi _ {\uparrow} ^{\rm IV} | {\sigma _0}{s_x} |  {\chi _ {\downarrow} ^{\rm IV}} \rangle }\\
{\langle \chi _ {\downarrow} ^{\rm IV} | {\sigma _0}{s_x} |  {\chi _ {\uparrow} ^{\rm IV}}   \rangle }&{\langle \chi _ {\downarrow} ^{\rm IV} | {\sigma _0}{s_x} |  {\chi _ {\downarrow} ^{\rm IV}} \rangle }
\end{array}} \right)\nonumber\\
& = & {\lambda}\frac{{{\rm{|}}\mu {\rm{|}}} -1}  {{{\rm{|1 + }}\mu {\rm{|}}}}\left( {\begin{array}{*{20}{c}}
0&1\\
1&0
\end{array}} \right)  ={m_{\rm IV}} \left( {\begin{array}{*{20}{c}}
0&1\\
1&0
\end{array}} \right).
\end{eqnarray}
Therefore, the four boundary effective masses of the diamond-shaped nanoflake can be expressed as $m_{\rm I} = m_{\rm II} = - m_{\rm III}=-m_{\rm IV}$.


\begin{thebibliography}{99}
\bibitem{Chiu2016} C.-K. Chiu, J. C. Y. Teo, A. P. Schnyder, and S. Ryu, Classification of Topological Quantum Matter with Symmetries, Rev. Mod. Phys. \textbf{88}, 035005 (2016).
\bibitem{Bansil2016} A. Bansil, H. Lin, and T. Das, Colloquium: Topological band theory, Rev. Mod. Phys. \textbf{88}, 021004 (2016).
\bibitem{Hasan2010} M. Z. Hasan and C. L. Kane, Colloquium: Topological insulators, Rev. Mod. Phys. \textbf{82}, 3045 (2010).

\bibitem{LiuFeng} F. Liu, Two-dimensional topological insulators: past, present and future, Coshare Science \textbf{01}, v3, 1-62 (2023).
\bibitem{Qi2011} X.-L. Qi and S.-C. Zhang, Topological insulators and superconductors, Rev. Mod. Phys. \textbf{83}, 1057 (2011).
\bibitem{Sato2010} M. Sato, Topological odd-parity superconductors, Phys. Rev. B \textbf{81}, 220504(R) (2010).

\bibitem{Armitage2018} N.P. Armitage, E. J. Mele, and A. Vishwanath, Weyl and Dirac semimetals in three-dimensional solids, Rev. Mod. Phys. \textbf{90}, 015001 (2018).
\bibitem{Lv2021} B. Q. Lv, T. Qian, and H. Ding, Experimental perspective on three-dimensional topological semimetals, Rev. Mod. Phys. \textbf{93}, 025002 (2021).
\bibitem{Burkov2016} A. A. Burkov, Topological semimetals, Nat. Mater. \textbf{15}, 1145 (2016).
\bibitem{Dai2016} X. Dai, Weyl fermions go into orbit, Nat. Phys. \textbf{12}, 727 (2016).


\bibitem{Bahari2019} M. Bahari and M. V. Hosseini, One-dimensional topological metal, Phys. Rev. B \textbf{99}, 155128 (2019).
\bibitem{Burkov2017} A. A. Burkov, Giant planar Hall effect in topological metals, Phys. Rev. B \textbf{96}, 041110(R) (2017).
\bibitem{Jangjan2021} M. Jangjan and M. V. Hosseini, Topological phase transition between a normal insulator and a topological metal state in a quasi-one-dimensional system, Sci. Rep. \textbf{11} 12966 (2021).
\bibitem{Xie2021} L. C. Xie, H. C. Wu, L. Jin, and Z. Song, Time-reversal symmetric topological metal, Phys. Rev. B \textbf{104}, 165422 (2021).
\bibitem{Zuo2023} Z.-W. Zuo, L. Lv, and D. Kang, Topological metals constructed by sliding quantum wire arrays, Phys. Rev. B \textbf{107}, 195142 (2023).
\bibitem{Yan2023} L. Yan, D. Zhang, X.-J. Wang and J.-Y. Yan, Intrinsic topological metal state in T-graphene, New J. Phys. \textbf{25} 043020 (2023).


\bibitem{Thouless1982} D. J. Thouless, M. Kohmoto, M. P. Nightingale, and M. den Nijs, Quantized Hall Conductance in a Two-Dimensional Periodic Potential, Phys. Rev. Lett. \textbf{49}, 405 (1982).
\bibitem{Xiao2010} D. Xiao, M.-C. Chang, and Q. Niu, Berry phase effects on electronic properties, Rev. Mod. Phys. \textbf{82}, 1959 (2010).
\bibitem{Teo2010} J. C. Y. Teo and C. L. Kane, Topological defects and gapless modes in insulators and superconductors, Phys. Rev. B \textbf{82}, 115120 (2010).


\bibitem{Haldane1988} F. D. M. Haldane, Model for a Quantum Hall Effect without Landau Levels: Condensed-Matter Realization of the "Parity Anomaly", Phys. Rev. Lett. \textbf{61}, 2015 (1988).
\bibitem{Kane2005} C. L. Kane and E. J. Mele, ${Z}_{2}$ Topological Order and the Quantum Spin Hall Effect, Phys. Rev. Lett. \textbf{95}, 146802 (2005).
\bibitem{Kane2005a} C. L. Kane and E. J. Mele, Quantum Spin Hall Effect in Graphene, Phys. Rev. Lett. \textbf{95}, 226801 (2005).
\bibitem{Bernevig2006} B. A. Bernevig, T. L. Hughes, and S.-C. Zhang, Quantum Spin Hall Effect and Topological Phase Transition in HgTe Quantum Wells, Science \textbf{314}, 1757 (2006).
\bibitem{Bernevig2006a} B. A. Bernevig and S.-C. Zhang, Quantum Spin Hall Effect, Phys. Rev. Lett. \textbf{96}, 106802 (2006).



\bibitem{Ying2018} X. Ying and A. Kamenev, Symmetry-Protected Topological Metals, Phys. Rev. Lett. \textbf{121}, 086810 (2018).
\bibitem{Ying2019} X. Ying and A. Kamenev, Topological transitions in metals, Phys. Rev. B \textbf{99}, 245411 (2019).
\bibitem{Xie2023} L. C. Xie, H. C. Wu, L. Jin, and Z. Song, Antihelical edge states in two-dimensional photonic topological metals, Sci. Bull. \textbf{68}, 255 (2023).
\bibitem{Dai2024} X. Dai and Q. Chen, Two-dimensional Weyl metal and second-order topological insulator phases in the modified Kane-Mele model, Phys. Rev. B \textbf{109}, 144108 (2024).



\bibitem{Benalcazar2017} W. A. Benalcazar, B. A. Bernevig, and T. L. Hughes, Quantized electric multipole insulators, Science \textbf{357}, 61 (2017).
\bibitem{Benalcazar2017a} W. A. Benalcazar, B. A. Bernevig, and T. L. Hughes, Electric multipole moments, topological multipole moment pumping, and chiral hinge states in crystalline insulators, Phys. Rev. B \textbf{96}, 245115 (2017).
\bibitem{Li2020} T. Li, P. Zhu, W. A. Benalcazar, and T. L. Hughes, Fractional disclination charge in two-dimensional ${C}_{n}$-symmetric topological crystalline insulators, Phys. Rev. B \textbf{101}, 115115 (2020).
\bibitem{Benalcazar2019} W. A. Benalcazar, T. Li, and T. L. Hughes, Quantization of fractional corner charge in ${C}_{n}$-symmetric higher-order topological crystalline insulators, Phys. Rev. B \textbf{99}, 245151 (2019).
\bibitem{Schindler2018} F. Schindler, A. M. Cook, M. G. Vergniory, Z. Wang, S. S. P. Parkin, B. A. Bernevig, and T. Neupert, Higher-order topological insulators, Sci. Adv. \textbf{4}, eaat0346 (2018).
\bibitem{Peterson2018} C. W. Peterson, W. A. Benalcazar, T. L. Hughes, and G. Bahl, A quantized microwave quadrupole insulator with topologically protected corner states, Nature \textbf{555}, 346 (2018).
\bibitem{Yao2023} J. Yao and L. Li, Extrinsic higher-order topological corner states in AB-stacked transition metal dichalcogenides, Phys. Rev. B \textbf{108}, 245131 (2023).
\bibitem{Bhowmik2024} S. Bhowmik, S. Banerjee, and A. Saha, Higher-order topological corner and bond-localized modes in magnonic insulators, Phys. Rev. B \textbf{109}, 104417 (2024).
\bibitem{Liu2023} Z.-R. Liu, C.-B. Hua, T. Peng, R. Chen, and B. Zhou, Higher-order topological insulators in hyperbolic lattices, Phys. Rev. B \textbf{107}, 125302 (2023).
\bibitem{Hung2024} Y.-C. Hung, B. Wang, C.-H. Hsu, A. Bansil, and H. Lin, Time-reversal soliton pairs in even spin Chern number higher-order topological insulators, Phys. Rev. B \textbf{110}, 035125 (2024).
\bibitem{Mazanov2024} M. Mazanov, A. S. Kupriianov, R. S. Savelev, Z. He, and M. A. Gorlach, Multipole higher-order topology in a multimode lattice, Phys. Rev. B \textbf{109}, L201122 (2024).


\bibitem{Ezawa2018} M. Ezawa, Minimal models for Wannier-type higher-order topological insulators and phosphorene, Phys. Rev. B \textbf{98}, 045125 (2018).
\bibitem{Park2019} M. J. Park, Y. Kim, G. Y. Cho, and S. Lee, Higher-Order Topological Insulator in Twisted Bilayer Graphene, Phys. Rev. Lett. \textbf{123}, 216803 (2019).
\bibitem{Liu2019} B. Liu, G. Zhao, Z. Liu, and Z. F. Wang, Two-Dimensional Quadrupole Topological Insulator in $\gamma$-Graphyne, Nano Lett. \textbf{19}, 6492 (2019).
\bibitem{Sheng2019} X.-L. Sheng, C. Chen, H. Liu, Z. Chen, Z.-M. Yu, Y. Zhao, and S. A. Yang, Two-Dimensional Second-Order Topological Insulator in Graphdiyne, Phys. Rev. Lett. \textbf{123}, 256402 (2019).



\bibitem{Ren2020} Y. Ren, Z. Qiao, and Q. Niu, Engineering Corner States from Two-Dimensional Topological Insulators, Phys. Rev. Lett. \textbf{124}, 166804 (2020).
\bibitem{Zhuang2022} Z.-Y. Zhuang and Z. Yan, Topological Phase Transitions and Evolution of Boundary States Induced by Zeeman Fields in Second-Order Topological Insulators, Front. Phys. \textbf{10}, 866347 (2022).
\bibitem{Han2022} B. Han, J. Zeng, and Z. Qiao, In-Plane Magnetization-Induced Corner States in Bismuthene, Chinese Phys. Lett. \textbf{39}, 017302 (2022).
\bibitem{Miao2022} C.-M. Miao, Q.-F. Sun, and Y.-T. Zhang,  Second-order topological corner states in zigzag graphene nanoflake with different types of edge magnetic configurations, Phys. Rev. B \textbf{106}, 165422 (2022).
\bibitem{Miao2023} C.-M. Miao, Y.-H. Wan, Q.-F. Sun, and Y.-T. Zhang, Engineering topologically protected zero-dimensional interface end states in antiferromagnetic heterojunction graphene nanoflakes, Phys. Rev. B \textbf{108}, 075401 (2023).
\bibitem{Miao2024} C.-M. Miao, L. Liu, Y.-H. Wan, Q.-F. Sun, and Y.-T. Zhang, General principle behind magnetization-induced second-order topological corner states in the Kane-Mele model, Phys. Rev. B \textbf{109}, 205417 (2024).
\bibitem{Chen2020a} C. Chen, Z. Song, J.-Z. Zhao, Z. Chen, Z.-M. Yu, X.-L. Sheng, and S. A. Yang, Universal Approach to Magnetic Second-Order Topological Insulator, Phys. Rev. Lett. \textbf{125}, 056402 (2020).
\bibitem{Li2024} Y.-X. Li, Y. Liu, and C.-C. Liu, Creation and manipulation of higher-order topological states by altermagnets, Phys. Rev. B \textbf{109}, L201109 (2024).
\bibitem{Liu2024} L. Liu, J. An, Y. Ren, Y.-T. Zhang, Z. Qiao, and Q. Niu, Engineering second-order topological insulators via coupling two first-order topological insulators, Phys. Rev. B \textbf{110}, 115427 (2024).

\bibitem{Lin2018} M. Lin and T. L. Hughes, Topological quadrupolar semimetals, Phys. Rev. B \textbf{98}, 241103(R) (2018).
\bibitem{Ghorashi2020} S. A. A. Ghorashi, T. Li, and T. L. Hughes, Higher-Order Weyl Semimetals, Phys. Rev. Lett. \textbf{125}, 266804 (2020).
\bibitem{Wang2020b} H.-X. Wang, Z.-K. Lin, B. Jiang, G.-Y. Guo, and J.-H. Jiang, Higher-Order Weyl Semimetals, Phys. Rev. Lett. \textbf{125}, 146401 (2020).
\bibitem{Xiong2023} Z. Xiong, Z.-K. Lin, H.-X. Wang, S. Liu, Y. Qian, and J.-H. Jiang, Valley higher-order Weyl semimetals, Phys. Rev. B \textbf{108}, 085141 (2023).
\bibitem{Pu2023} Z. Pu, H. He, L. Luo, Q. Ma, L. Ye, M. Ke, and Z. Liu, Acoustic Higher-Order Weyl Semimetal with Bound Hinge States in the Continuum, Phys. Rev. Lett. \textbf{130}, 116103 (2023).

\bibitem{Ghorashi2021} S. A. A. Ghorashi, T. Li, M. Sato, and T. L. Hughes, Non-Hermitian higher-order Dirac semimetals, Phys. Rev. B \textbf{104}, L161116 (2021).
\bibitem{Wang2022} Z. Wang, D. Liu, H. T. Teo, Q. Wang, H. Xue, and B. Zhang, Higher-order Dirac semimetal in a photonic crystal, Phys. Rev. B \textbf{105}, L060101 (2022).
\bibitem{Chen2022} C. Chen, X.-T. Zeng, Z. Chen, Y. X. Zhao, X.-L. Sheng, and S. A. Yang, Second-Order Real Nodal-Line Semimetal in Three-Dimensional Graphdiyne, Phys. Rev. Lett. \textbf{128}, 026405 (2022).
\bibitem{Gao2023} M.-J. Gao, H. Wu, and J.-H. An, Engineering second-order nodal-line semimetals by breaking $\mathcal{PT}$ symmetry and periodic driving, Phys. Rev. B \textbf{107}, 035128 (2023).
\bibitem{Du2022} X.-L. Du, R. Chen, R. Wang, and D.-H. Xu, Weyl nodes with higher-order topology in an optically driven nodal-line semimetal, Phys. Rev. B \textbf{105}, L081102 (2022).
\bibitem{Hirsbrunner2024} M. R. Hirsbrunner, A. D. Gray, and T. L. Hughes, Crystalline electromagnetic responses of higher-order topological semimetals, Phys. Rev. B \textbf{109}, 075169 (2024).
\bibitem{Pan2024} B. Pan, Y. Hu, P. Zhou, H. Xiao, X. Yang, and L. Sun, Higher-order double-Weyl semimetal, Phys. Rev. B \textbf{109}, 035148 (2024).
\bibitem{Qi2024} Y. Qi, Z. He, K. Deng, J. Li, and Y. Wang, Multipole higher-order topological semimetals, Phys. Rev. B \textbf{109}, L060101 (2024).



\bibitem{Gmitra2015} M. Gmitra and J. Fabian, Graphene on transition-metal dichalcogenides: A platform for proximity spin-orbit physics and optospintronics, Phys. Rev. B \textbf{92}, 155403 (2015).
 \bibitem{Ghiasi2017} T. S. Ghiasi, J. Ingla-Ayn$\acute{\rm e}$s, A. A. Kaverzin, and B. J. van Wees, Large Proximity-Induced Spin Lifetime Anisotropy in Transition-Metal Dichalcogenide/Graphene Heterostructures, Nano Lett. \textbf{17}, 7528 (2017).
 \bibitem{Zihlmann2018} S. Zihlmann, A.W. Cummings, J. H. Garcia, M. Kedves, K. Watanabe, T. Taniguchi, C. Sch$\ddot{\rm o}$nenberger, and P. Makk, Large spin relaxation anisotropy and valley-Zeeman spin-orbit coupling in ${\mathrm{WSe}}_{2}$/graphene/$h$-BN heterostructures, Phys. Rev. B \textbf{97}, 075434 (2018).
  \bibitem{Wang2015} Z. Wang, D.-K. Ki, H. Chen, H. Berger, A. H. MacDonald, and A. F. Morpurgo, Strong interface-induced spin-orbit interaction in graphene on WS$_2$, Nat. Commun. \textbf{6}, 8339 (2015).
 \bibitem{Benitez2018} L. A. Ben$\acute{\rm l}$tez, J. F. Sierra, W. S. Torres, A. Arrighi, F. Bonell, M. V. Costache, and S. O. Valenzuela, Strongly anisotropic spin relaxation in graphene-transition metal dichalcogenide heterostructures at room temperature, Nat. Phys. \textbf{14}, 303 (2018).

  \bibitem{Colomes2018} E. Colom$\acute{\rm e}$s and M. Franz, Antichiral Edge States in a Modified Haldane Nanoribbon, Phys. Rev. Lett. \textbf{120}, 086603 (2018).

 {\color{blue}\bibitem{Frank2018} T. Frank, P. H$\rm \ddot{o}$gl, M. Gmitra, D. Kochan, and J. Fabian, Protected pseudohelical edge states in ${\mathbb{Z}}_{2}$-trivial proximitized graphene, Phys. Rev. Lett. \textbf{120}, 156402 (2018).}

\bibitem{Fisher1981} D. S. Fisher and P. A. Lee, Relation between conductivity and transmission matrix, Phys.Rev.B \textbf{23}, 6851 (1981).

\bibitem{Metalidis2005} G. Metalidis and P. Bruno, Green's function technique for studying electron flow in two-dimensional mesoscopic samples, Phys.Rev.B \textbf{72}, 235304 (2005).

\bibitem{Sun2009} Q.-F. Sun and X. C. Xie, Quantum transport through a graphene nanoribbon-superconductor junction, J. Phys.: Condens. Matter \textbf{21}, 344204 (2009).
\bibitem{Lee1981} D. H. Lee and J. D. Joannopoulos, Simple scheme for surface-band calculations. II. The Green's function, Phys. Rev. B \textbf{23}, 4997 (1981).


\bibitem{Jackiw1976} R. Jackiw and C. Rebbi, Solitons with fermion number 1/2, Phys. Rev. D \textbf{13}, 3398 (1976).



\bibitem{Cummings2017} A. W. Cummings, J. H. Garcia, J. Fabian, and S. Roche, Giant Spin Lifetime Anisotropy in Graphene Induced by Proximity Effects, Phys. Rev. Lett. \textbf{119}, 206601 (2017).

\bibitem{Omar2017} S. Omar and B. J. van Wees, Graphene-${\mathrm{WS}}_{2}$ heterostructures for tunable spin injection and spin transport, Phys. Rev. B \textbf{95}, 081404(R) (2017).
\bibitem{Avsar2014} A. Avsar, J. Y. Tan, T. Taychatanapat, J. Balakrishnan, G. K. W. Koon, Y. Yeo, J. Lahiri, A. Carvalho, A. S. Rodin, E. C. T. O'Farrell, G. Eda, A. H. Castro Neto, and B. $\rm{\ddot{O}}$zyilmaz, Spin-Orbit proximity effect in graphene, Nat. Commun. \textbf{5}, 4875 (2014).



\bibitem{Volkl2017} T. V$\rm{\ddot{o}}$lkl, T. Rockinger, M. Drienovsky, K. Watanabe, T. Taniguchi, D.Weiss, and J. Eroms, Magnetotransport in heterostructures of transition metal dichalcogenides and graphene, Phys. Rev. B \textbf{96}, 125405 (2017).
\bibitem{Wang2016} Z. Wang, D.-K. Ki, J. Y. Khoo, D. Mauro, H. Berger, L. S. Levitov, and A. F. Morpurgo, Origin and Magnitude of `Designer' Spin-Orbit Interaction in Graphene on Semiconducting Transition Metal Dichalcogenides, Phys. Rev. X \textbf{6}, 041020 (2016).


{\color{blue}

\bibitem{Britnell2013} L. Britnell, R. M. Ribeiro, A. Eckmann, R. Jalil, B. D. Belle, A. Mishchenko, Y. J. Kim, R. V. Gorbachev, T. Georgiou, S. V. Morozov, A. N. Grigorenko, A. K. Geim, C. Casiraghi, A. H. C. Neto, K. S. Novoselov, Strong light-matter interactions in heterostructures of atomically thin films, Science \textbf{340}, 1311 (2013).
\bibitem{Fang2015}  W. Fang, H. Zhao, Y. Xie, J. Fang, J. Xu, Z. Chen, Facile hydrothermal synthesis of VS2/Graphene nanocomposites with superior high-rate capability as lithium-ion battery cathodes, ACS Appl. Mater. Interfaces \textbf{7}, 13044 (2015).
\bibitem{Li2018}  C. Li, Q. Cao, F. Wang, Y. Xiao, Y. Li, J.-J. Delaunay, H. Zhu, Engineering graphene and TMDs based van der Waals heterostructures for photovoltaic and photoelectrochemical solar energy conversion, Chem. Soc. Rev. \textbf{47}, 4981 (2018).
\bibitem{Wang2017} T. Wang, C. Liu, F. Jiang, Z. Xu, X. Wang, X. Li, C. Li, J. Xu, X. Yang, Solution-processed two-dimensional layered heterostructure thin-film with optimized thermoelectric performance, J. Chem. Soc. Faraday Trans. \textbf{19}, 17560 (2017).
}

\bibitem{Sheng2017} X.-L. Sheng and B. K. Nikoli$\rm{\acute{c}}$, Monolayer of the $5d$ transition metal trichloride ${\mathrm{OsCl}}_{3}$: A playground for two-dimensional magnetism, room-temperature quantum anomalous Hall effect, and topological phase transitions, Phys. Rev. B \textbf{95}, 201402(R) (2017).
\bibitem{Liu2018} Z. Liu, G. Zhao, B. Liu, Z. F. Wang, J. Yang, and F. Liu, Intrinsic Quantum Anomalous Hall Effect with In-Plane Magnetization: Searching Rule and Material Prediction, Phys. Rev. Lett. \textbf{121}, 246401 (2018).
\bibitem{Webster2018} L. Webster and J.-A. Yan, Strain-tunable magnetic anisotropy in monolayer ${\mathrm{CrCl}}_{3}$, ${\mathrm{CrBr}}_{3}$, and ${\mathrm{CrI}}_{3}$, Phys. Rev. B \textbf{98}, 144411 (2018).

\bibitem{Cai2019} X. Cai, T. Song, N. P. Wilson, G. Clark, M. He, X. Zhang, T. Taniguchi, K. Watanabe, W. Yao, D. Xiao, M. A. McGuire, D. H. Cobden, and X. Xu, Atomically thin $\rm{CrCl_3}$: An in-plane layered antiferromagnetic insulator, Nano Lett. \textbf{19}, 3993 (2019).
\bibitem{Umemoto2019} Y. Umemoto, K. Sugawara, Y. Nakata, T. Takahashi, and T. Sato, Pseudogap, Fermi arc, and Peierls-insulating phase induced by 3D-2D crossover in monolayer VSe$_2$, Nano Res. \textbf{12}, 165 (2019).
\bibitem{Bonilla2018} M. Bonilla, S. Kolekar, Y. Ma, H. C. Diaz, V. Kalappattil, R. Das, T. Eggers, H. R. Gutierrez, M.-H. Phan, and M. Batzill, Strong room-temperature ferromagnetism in VSe$_2$ monolayers on van der Waals substrates, Nat. Nanotechnol. \textbf{13}, 289 (2018).
\bibitem{Jenjeti2018} R. N. Jenjeti, R. Kumar, M. P. Austeria, and S. Sampath, Field Effect Transistor Based on Layered NiPS$_3$, Sci. Rep. \textbf{8}, 8586 (2018).
\bibitem{Kim2018} S. Y. Kim, T. Y. Kim, L. J. Sandilands, S. Sinn, M.-C. Lee, J. Son, S. Lee, K.-Y. Choi, W. Kim, B.-G. Park, C. Jeon, H. D. Kim, C.-H. Park, J.-G. Park, S. J. Moon, and T. W. Noh, Charge-Spin Correlation in Van Der Waals Antiferromagnet $\rm{NiPS_3}$, Phys. Rev. Lett. \textbf{120}, 136402 (2018).
\bibitem{Gong2019} C. Gong and X. Zhang, Two-dimensional magnetic crystals and emergent heterostructure devices, Science \textbf{363}, eaav4450 (2019).

\bibitem{Harper1955} P. G. Harper, Single Band Motion of Conduction Electrons in a Uniform Magnetic Field, Proc. Phys. Soc. Lond. A \textbf{68}, 874 (1955).

\end{thebibliography}
\end{document}